# Universal *T*-linear resistivity and Planckian limit in overdoped cuprates

A. Legros[1,2], S. Benhabib[3], W. Tabis[3,4], F. Laliberté[1], M. Dion[1], M. Lizaire[1],

B. Vignolle[3], D. Vignolles[3], H. Raffy[5], Z. Z. Li[5], P. Auban-Senzier[5],

N. Doiron-Leyraud[1], P. Fournier[1,6], D. Colson[2], L. Taillefer[1,6], and C. Proust[3,6]

*1 Institut quantique, Département de physique & RQMP, Université de Sherbrooke, Sherbrooke, Québec J1K 2R1, Canada*

*2 Service de Physique de l'État Condensé (CEA, CNRS), Université Paris-Saclay, CEA Saclay, Gif-sur-Yvette 91191, France*

*3 Laboratoire National des Champs Magnétiques Intenses (CNRS, EMFL, INSA, UJF, UPS), Toulouse 31400, France*

*4 AGH University of Science and Technology, Faculty of Physics and Applied Computer Science, Al. Mickiewicza 30, 30-059 Krakow, Poland*

*5 Laboratoire de Physique des Solides, Université Paris-Sud, Université Paris-Saclay, CNRS UMR 8502, Orsay 91405, France*

*6 Canadian Institute for Advanced Research, Toronto, Ontario M5G 1Z8, Canada*

The perfectly linear temperature dependence of the electrical resistivity observed as $T \rightarrow 0$ in a variety of metals close to a quantum critical point[1,2,3,4] is a major puzzle of condensed matter physics[5]. Here we show that *T*-linear resistivity as $T \rightarrow 0$ is a generic property of cuprates, associated with a universal scattering rate. We measured the low-temperature resistivity of the bi-layer cuprate $Bi_2Sr_2CaCu_2O_{8+\delta}$ and found that it exhibits a *T*-linear dependence with the same slope as in the single-layer cuprates $Bi_2Sr_2CuO_{6+\delta}$ (ref. 6), $La_{1.6-x}Nd_{0.4}Sr_xCuO_4$ (ref. 7) and $La_{2-x}Sr_xCuO_4$ (ref. 8), despite their very different Fermi surfaces and structural, superconducting and magnetic properties. We then show that the *T*-linear coefficient (per $CuO_2$ plane), $A_1^\square$, is given by the universal relation $A_1^\square T_F = h / 2e^2$, where $e$ is the electron charge, $h$ is the Planck constant and $T_F$ is the Fermi temperature. This relation, obtained by assuming that the scattering rate $1 / \tau$ of charge carriers reaches the Planckian limit[9,10], whereby $\hbar / \tau = k_B T$, works not only for hole-doped cuprates[6,7,8,11,12] but also for electron-doped cuprates[13,14], despite the different nature of their quantum critical point and strength of their electron correlations.

In conventional metals, the electrical resistivity $\rho(T)$ normally varies as $T^2$ in the limit $T \rightarrow 0$, where electron-electron scattering dominates, in accordance with Fermi-liquid theory. However, close to a quantum critical point (QCP) where a phase of antiferromagnetic (AF) order ends, $\rho(T) \sim T^n$, with $n < 2.0$. Most striking is the observation of a perfectly linear $T$ dependence $\rho(T) = \rho_0 + A_1 T$ as $T \rightarrow 0$ in several very different materials, when tuned to their magnetic QCP, for example: the quasi-1D organic conductor $(TMTSF)_2PF_6$ (ref. 4); the quasi-2D ruthenate $Sr_3Ru_2O_7$ (ref. 3); and the 3D heavy-fermion metal $CeCu_6$ (ref. 1). This $T$-linear resistivity as $T \rightarrow 0$ has emerged as one of the major puzzles in the physics of metals[5], and while several theoretical scenarios have been proposed[15], no compelling explanation has been found.

In cuprates, a perfect $T$-linear resistivity as $T \rightarrow 0$ has been observed (once superconductivity is suppressed by a magnetic field) in two closely related electron-doped materials, $Pr_{2-x}Ce_xCuO_4$ (PCCO) (refs. 2,16,17) and $La_{2-x}Ce_xCuO_4$ (LCCO) (refs. 13,14), and in three hole-doped materials: $Bi_2Sr_2CuO_{6+\delta}$ (ref. 6), $La_{2-x}Sr_xCuO_4$ (LSCO) (ref. 8) and $La_{1.6-x}Nd_{0.4}Sr_xCuO_4$ (Nd-LSCO) (refs. 7,11,12). On the electron-doped side, $T$-linear resistivity is seen just above the QCP (ref. 16) where AF order ends[18] as a function of $x$, and as such it may not come as a surprise. On the hole-doped side, however, the doping values where $\rho(T) = \rho_0 + A_1 T$ as $T \rightarrow 0$ are very far from the QCP where long-range AF order ends ($p_N \sim 0.02$), e.g. at $p = 0.24$ in Nd-LSCO (Fig. 1a) and in the range $p = 0.21$-$0.26$ in LSCO (Fig. 1b). Instead, these values are close to the critical doping where the pseudogap phase ends, i.e. at $p^* = 0.23 \pm 0.01$ in Nd-LSCO (ref. 11) and at $p^* \sim 0.18$-$0.19$ in LSCO (ref. 8), where the role of AF spin fluctuations is not clear. In Bi2201, $p^*$ is further still (see Supplementary Section 10).

To make progress, several questions must be answered. Is $T$-linear resistivity as $T \rightarrow 0$ in hole-doped cuprates limited to single-layer materials with low $T_c$, or is it generic ? Why is $\rho(T) = \rho_0 + A_1 T$ as $T \rightarrow 0$ seen in LSCO over an anomalously wide doping range[8] ? Is there a common mechanism linking cuprates to the other metals where $\rho \sim T$ as $T \rightarrow 0$ ?

To establish the universal character of $T$-linear resistivity in cuprates, we have turned to $Bi_2Sr_2CaCu_2O_{8+\delta}$ (Bi2212). While Nd-LSCO and LSCO have essentially the same single electron-like diamond-shaped Fermi surface at $p > p^*$ (refs. 19,20), Bi2212 has a very different Fermi surface, consisting of two sheets, one of which is also diamond-like at $p > 0.22$, but the other is much more circular[21] (see Supplementary Section 1). Moreover, the structural, magnetic, and superconducting properties of Bi2212 are very different to those of Nd-LSCO and LSCO: a stronger 2D character, a larger gap to spin excitations, no spin-density-wave order above $p \sim 0.1$, a much higher superconducting $T_c$ .

We measured the resistivity of Bi2212 at $p = 0.23$ by suppressing superconductivity with a magnetic field of 58 T. At $p = 0.23$, the system is just above its pseudogap critical point ($p^* = 0.22$ (ref. 22); see Supplementary Section 2). Our data are shown in Fig. 2. The raw data at $H = 55$ T reveal a perfectly linear $T$ dependence of $\rho(T)$ down to the lowest accessible temperature (Fig. 1a). Correcting for the magneto-resistance (see Methods and Supplementary Section 3), as was done for LSCO (ref. 8), we find that the $T$-linear dependence of $\rho(T)$ seen in Bi2212 at $H = 0$ from $T \sim 120$ K down to $T_c$ simply continues to low temperature, with the same slope $A_1 = 0.62 \pm 0.06$ $\mu\Omega$ cm / K (Fig. 2b). Measured per $CuO_2$ plane, this gives $A_1^\square \equiv A_1 / d = 8.0 \pm 0.9$ $\Omega$ / K, where $d$ is the (average) separation between $CuO_2$ planes. Remarkably, this is the same value, within error bars, as measured in Nd-LSCO at $p = 0.24$, where $A_1^\square = 7.4 \pm 0.8$ $\Omega$ / K (see Table 1).

The observation of $T$-linear resistivity in those two cuprates shows that it is robust against changes in the shape, topology and multiplicity of the Fermi surface. By contrast, the Hall coefficient $R_H$ is not. In Fig. 2d, we compare $R_H(T)$ in Bi2212 and in Nd-LSCO (and PCCO). We see strong differences, brought about by the different anisotropies in either the inelastic scattering or the Fermi surface, or both[23]. Nevertheless, $\rho(T)$ is perfectly linear in both cases. Moreover, the coefficient $A_1^\square$ is the same despite the very different spectra of low-energy spin fluctuations, gapped in Bi2212 (ref. 24) and ungapped in Nd-LSCO (ref. 25). We conclude that a $T$-linear resistivity as $T \to 0$ is a generic and robust property of cuprates.

(Note that $\rho(T)$ deviates from pure $T$-linearity above a certain temperature, and in this high-$T$ regime a generic evolution has also been found[26], with $\rho(T) \sim A_1 T + A_2 T^2$. Here we focus strictly on the low-$T$ regime of pure $T$-linear resistivity.)

We now investigate the strength of this $T$-linear resistivity, *i.e.* the magnitude of $A_1$. In Fig. 3b, we plot $A_1^\square$ vs $p$ for hole-doped cuprates. We see from the LSCO data[8] that $A_1^\square$ increases with decreasing $p$ (Fig. 1b), from $A_1^\square \sim 8$ $\Omega$ / K at $p = 0.26$ to $A_1^\square \sim 15$ $\Omega$ / K at $p = 0.21$ (see Table 3, Methods). In Nd-LSCO, we see a similar increase (Figs. 1c and 3b), when pressure[12] is used to suppress the onset of the pseudogap at $p = 0.22$ and $p = 0.23$ (see Supplementary Section 4). In Fig. 1d, we present our data on PCCO at $x = 0.17$ (see also Supplementary Section 5), and compare with prior data on LCCO (ref. 14; Supplementary Section 6). In Fig. 4b, we plot $A_1^\square$ vs $x$ for electron-doped cuprates, and see that $A_1^\square$ also increases with decreasing $x$, from $A_1^\square \sim 1.5$ $\Omega$ / K at $x = 0.17$ to $A_1^\square \sim 3$ $\Omega$ / K at $x = 0.15$ (see Table 5, Methods). Note that these values are 5 times smaller than in hole-doped cuprates.

To summarize : i) $A_1^\square$ increases as the doping is reduced in both hole-doped and electron-doped cuprates; ii) $A_1^\square$ is much larger in hole-doped cuprates; iii) $T$-linear resistivity as $T \rightarrow 0$ is observed over a range of doping, not just at one doping; iv) $T$-linear resistivity does not depend on the nature of the inelastic scattering process (hole-doped vs electron-doped) or on the topology of the Fermi surface (LSCO vs NCCO, Bi2212 vs Nd-LSCO; Supplementary Section 1).

To explain these experimental facts, we consider the empirical observation that the strength of the $T$-linear resistivity for several metals is approximately given by a scattering rate that has a universal value, namely $\hbar / \tau = k_B T$ (ref. 10), and test it in cuprates. This observation suggests that a $T$-linear regime will be observed whenever $1 / \tau$ reaches its Planckian limit, $k_B T / \hbar$, irrespective of the underlying mechanism for inelastic scattering[9]. In the simple case of an isotropic Fermi surface, the connection between $\rho$ and $\tau$ is given by the Drude formula, $\rho = (m^* / n\, e^2)\, (1 / \tau)$, where $n$ is the carrier density and $m^*$ is the effective mass. So when $\rho(T) = \rho_0 + A_1 T$, then $A_1 = (m^* / n\, e^2)\, (1/\tau)\, (1/T) = \alpha\, (\, m^* / n\, )\, (k_B / e^2\, \hbar\, )$, with $\hbar / \tau \equiv \alpha\, k_B T$. In 2D, this can be written as :

$$A_1^\square\ =\ \alpha\, (\, h / 2e^2\, )\ \ 1 / T_F\ ,\qquad\qquad (1)$$

where $T_F\ =\ (\pi\, \hbar^2 / k_B)\, (n\, d / m^*)$ is the Fermi temperature.

Let us first evaluate $\alpha$ in electron-doped cuprates, where the Drude formula is expected to work well, since their single Fermi surface is highly 2D and circular (in the overdoped region[27]; see Supplementary Section 1). Quantum oscillations in $Nd_{2-x}Ce_xCuO_4$ (NCCO) provide a direct and precise measurement of $n$ and $m^*$ in electron-doped cuprates[28,29]. The Luttinger rule sets the carrier density to be $n = (1-x) / (a^2\, d)$, given precisely by the oscillation frequency $F = n\, d\, (h/2e)$, where $x$ is the number of doped electrons per Cu atom and $a$ is the in-plane lattice constant. In Fig. 4a, we see that $m^*$ increases from 2.3 $m_0$ at $x = 0.175$ to 3.0 $m_0$ at $x = 0.151$, where $m_0$ is the bare electron mass (Table 4, Methods). This increasing value is consistent, within error bars, with specific heat data in PCCO at $x = 0.15$, where $\gamma = 5.5 \pm 0.3$ mJ / $K^2$ mol (ref. 30), which yields $m^* = 3.6 \pm 0.3$ $m_0$ (see Eq. 2 below). We use $n$ and $m^*$ to estimate $T_F$ and then plot, in Fig. 4b, the value of $A_1^\square$ predicted by Eq. 1, for $\alpha = 1$ (solid line in Fig. 4b; Table 4, Methods). Comparison with the measured values of $A_1^\square$ in PCCO (red hexagon in Fig. 4b) and in LCCO (blue circles in Fig. 4b), listed in Table 5 of Methods, shows that the scattering rate in electron-doped cuprates is given by $\hbar / \tau = \alpha\, k_B\, T$, with $\alpha = 1.0 \pm 0.3$, $i.e.$ the Planckian limit is observed, within experimental error bars.

Let us now turn to hole-doped cuprates. Here our quantitative estimates will be more approximate, since Fermi surfaces are not circular but diamond-shaped (Supplementary Section 1), but we are looking for a large effect (factor ~5 in $A_1^\square$ relative to electron-doped materials) and a qualitative trend (increase in $A_1^\square$ as $p$ is reduced towards $p^*$). In the absence of quantum oscillation data for Bi2212, LSCO, Nd-LSCO and Bi2201, we estimate $m^*$ from specific heat data, since in 2D the specific heat coefficient $\gamma$ is directly related to $m^*$ :

$$\gamma = (\pi N_A k_B^2 / 3\hbar^2) \ a^2 \ m^* \ . \qquad (2)$$

for a single Fermi surface, where $N_A$ is Avogadro's number. This connection between $m^*$ and $\gamma$ was nicely confirmed by quantum oscillations in $Tl_2Ba_2CuO_{6+\delta}$ at $p \sim 0.3$, where $m^* = 5.2 \pm 0.4 \ m_0$ and $\gamma = 7 \pm 1$ mJ / $K^2$ mol (ref. 31). In Bi2212, $\gamma = 12 \pm 2$ mJ / $K^2$ mol-Cu at $p = 0.22 = p^*$ (ref. 32; see Supplementary Section 8), giving $m^* = 8.4 \pm 1.6 \ m_0$ (Eq. 2). Applying Eq. 1, with $n (a^2 d) = 1 - p = 0.77$ (for an electron-like Fermi surface; Supplementary Section 1), the Planckian limit predicts $A_1^\square = 7.4 \pm 1.4 \ \Omega$ / K, while we measured $A_1^\square = 8.0 \pm 0.9 \ \Omega$ / K, so that $\alpha = 1.1 \pm 0.3$.

In LSCO, $\gamma$ increases from $\gamma = 6.9 \pm 1$ mJ / $K^2$ mol at $p = 0.33$ (ref. 33) to $\gamma = 14 \pm 2$ mJ / $K^2$ mol at $p = 0.26$ (ref. 34), showing that $m^*$ increases with reduced doping also in hole-doped cuprates (solid line in Fig. 3a). Applying Eq. 1 to LSCO data at $p = 0.26$, using $n (a^2 d) = 1 - p = 0.74$ and $m^* = 9.8 \pm 1.7 \ m_0$ (Eq. 2; Table 2 in Methods), the Planckian limit predicts $A_1^\square = 8.9 \pm 1.8 \ \Omega$ / K, while we see $A_1^\square = 8.2 \pm 1.0 \ \Omega$ / K (Fig. 1b; Table 3 in Methods), so that $\alpha = 0.9 \pm 0.3$.

In Nd-LSCO, an increase in $m^*$ has also been observed in recent specific heat measurements[35], from $\gamma = 5.4 \pm 1$ mJ / $K^2$ mol at $p = 0.40$ to $\gamma = 11 \pm 1$ mJ / $K^2$ mol at $p = 0.27$ (Fig. 3a). At $p = 0.24$, the electronic specific heat $C_{el}$ varies as $C_{el} / T \sim \log(1/T)$, which complicates the estimation of $m^*$. Taking the mean value between $C_{el} / T = 12$ mJ / $K^2$ mol at 10 K and $C_{el} / T = 22$ mJ / $K^2$ mol at 0.5 K (ref. 35), we get $m^* = 12 \pm 4 \ m_0$ and hence $\alpha = 0.7 \pm 0.4$, consistent with the Planckian limit for a third hole-doped material. See Table 1 for a summary of the numbers.

Finally, a stringent test of whether the Planckian limit operates in cuprates is provided by Bi2201, since in this particular cuprate the pseudogap critical point that controls $T$-linear scattering occurs at a much higher doping than in other cuprates, namely $p^* \sim 0.4$ (see Supplementary Section 10). Despite this doubling of $p^*$ and the very different volume of the Fermi surface relative to Bi2212, LSCO and Nd-LSCO, we find that $\alpha = 1.0 \pm 0.4$ in Bi2201 (Table 1; Supplementary Section 10).

In summary, our estimations reveal that the scattering rate responsible for the $T$-linear resistivity in PCCO, LCCO, Bi2212, LSCO, Nd-LSCO and Bi2201 tends to the same universal value, namely $\hbar / \tau = \alpha \ k_B \ T$, with $\alpha = 1.0$ (Table 1). A constant value of $\alpha$ in Eq. 1 implies that $A_1^\square \sim 1 / T_F$, so that, in essence, $A_1^\square \sim m^*$. This explains why the slope of the $T$-linear resistivity is much larger in hole-doped than in electron-doped cuprates, since the effective mass is much higher in the former (Fig. 3a vs Fig. 4a). It also explains why $A_1^{\square}$ increases in LSCO when going from $p = 0.26$ to $p = 0.21$ (Fig. 1b) and in Nd-LSCO (under pressure) when going from $p = 0.24$ to $p = 0.22$ (Fig. 1c). Indeed, as shown in Fig. 3, $A_1^{\square}$ (Fig. 3b) and $m^*$ (Fig. 3a) in LSCO and Nd-LSCO are seen to rise in tandem with decreasing $p$ (we make the natural assumption that $m^*$ continues to rise until $p$ reaches $p^*$). Moreover, a Planckian limit on scattering provides an explanation for the "anomalous" range in doping over which $\rho \sim A_1 T$ is observed in LSCO (ref. 8). As doping decreases below $p \sim 0.33$, scattering increases steadily until $p^* \sim 0.18\text{-}19$, but the inelastic scattering rate $1/\tau$ cannot exceed the Planckian limit, reached at $p \sim 0.26$. So between $p \sim 0.26$ and $p = p^*$, $\rho(T)$ is linear and $1/\tau$ saturates. The continuous increase of $A_1$ below $p \sim 0.26$ can be understood if we assume that $m^*$ continues to increase in the range $p^* < p < 0.26$ (ref. 35), since $A_1 \sim m^* (1/\tau) \sim m^*$. If one could lower $p^*$, the range of $T$-linear resistivity would expand further. This is indeed what happens in Nd-LSCO when $p^*$ is lowered by applying pressure[12] (Fig. 3b).

The fact that $\alpha \sim 1.0$ in cuprates has far-reaching implications since other metals with $T$-linear resistivity as $T \to 0$ also appear to have $\alpha \sim 1.0$ (ref. 10). The case is particularly clear in the organic conductor (TMTSF)$_2$PF$_6$, a well-characterized single-band metal whose resistivity is perfectly $T$-linear as $T \to 0$ (ref. 4), where $\alpha = 1.0 \pm 0.3$ (see Supplementary Section 9). For such dramatically different metals as the quasi-1D organics and the cuprates – not to mention the heavy-fermion metals and the pnictides[10] – to all have quantitatively the same scattering rate in their respective $T$-linear regimes, there must be a fundamental and universal principle at play. Our findings support the idea[9,10] that $T$-linear resistivity is achieved when the scattering rate hits the Planckian limit, given by $\hbar/\tau = k_B T$, whatever the scattering process, whether by AF spin fluctuations or not. If Planckian dissipation is the fundamental principle, new theoretical approaches are needed to understand how it works[36,37,38].

**Acknowledgements** A portion of this work was performed at the LNCMI, a member of the European Magnetic Field Laboratory (EMFL). C.P. acknowledges funding from the French ANR SUPERFIELD, and the LABEX NEXT. P.F. and L.T. acknowledge support from the Canadian Institute for Advanced Research (CIFAR) and funding from the Natural Sciences and Engineering Research Council of Canada (NSERC), the Fonds de recherche du Québec - Nature et Technologies (FRQNT), and the Canada Foundation for Innovation (CFI). L.T. acknowledges support from a Canada Research Chair. This research was undertaken thanks in part to funding from the Canada First Research Excellence Fund. Part of this work was funded by the Gordon and Betty Moore Foundation's EPiQS Initiative (Grant GBMF5306 to L.T.).

# METHODS

## SAMPLES

<u>Bi2212.</u> Our thin film of $Bi_2Sr_2CaCu_2O_8$ (Bi2212) was grown epitaxially at 740°C on a $SrTiO_3$ substrate by rf-magnetron sputtering with $O_2$/Ar gas and fully oxygen overdoped after deposition (see ref. 12 in ref. 21). The film thickness was measured by deposition rate calibration, giving $t$ = 240 ± 15 nm. The film was patterned by mechanical scribing (avoiding lithography resist) into the shape of a Hall bar consisting of two large pads (for current) connected by a narrow bridge (275 µm wide) between 2 couples of voltage pads distant by 1.15 mm for longitudinal and transverse resistance measurements. Six gold contacts were deposited by sputtering on the different pads and gold wires were attached with silver paint.

The superconducting transition temperature $T_c$ = 50 K was determined as the temperature below which the zero-field resistance $R$ = 0. The hole doping $p$ is obtained from $T_c$, using the usual convention[22,32], according to which our overdoped sample has a nominal doping $p$ = 0.23. This means that its doping is just slightly above the end of the pseudogap phase[22] (see Supplementary Section 2). It is also just above the Lifshitz transition where its anti-bonding band crosses the Fermi level to produce an electron-like diamond-shaped Fermi surface[21] (see Supplementary Section 1).

<u>PCCO.</u> Our thin films of $Pr_{2-x}Ce_xCuO_4$ (PCCO) were grown by pulsed laser deposition on LSAT substrates under 200mTorr of $N_2O$ using targets including an excess of Cu to suppress the growth of parasitic phases[39]. Films were then annealed 4 minutes in vacuum. The film thickness was measured via the width of x-ray diffraction peaks, giving $t$ = 230 ± 30 nm. A very small amount of parasitic phase was detected in the XRD spectra. However, its impact on the cross section of the films should be much smaller than the uncertainty coming from the thickness measurement. Six indium-silver contacts were applied in the standard geometry.

The superconducting transition temperature $T_c$ = 13 K was determined as the temperature below which the zero-field resistance $R$ = 0. The electron concentration is taken to be the cerium content, $x$ = 0.17, with an error bar ± 0.005. This means that our samples have a concentration slightly above the quantum critical point where the Fermi surface of PCCO is known to undergo a reconstruction by AF ordering[16]. The Fermi surface of NCCO at that doping could not be simpler: it is a single circular cylinder[27] (see Supplementary Section 1).

## MEASUREMENT OF THE LONGITUDINAL AND TRANSVERSE RESISTANCES

The longitudinal resistance $R_{xx}$ and transverse (Hall) resistance $R_{xy}$ of our Bi2212 film were measured in Toulouse in pulsed fields up to 58 T. The measurements were performed using a conventional 6-point configuration with a current excitation of 0.5 mA at a frequency of ~ 10 kHz. A high-speed acquisition system was used to digitize the reference signal (current) and the voltage drop across the sample at a frequency of 500 kHz. The data were post-analyzed with a software to perform the phase comparison. Data for the rise and fall of the field pulse were in good agreement, thus excluding any heating due to eddy currents. Tests at different frequencies showed excellent reproducibility.

$R_{xx}$ and $R_{xy}$ of our Bi2212 film were also measured in Orsay, at $H$ = 0 and $H$ = 9 T, respectively.

The longitudinal resistance $R_{xx}$ of our three PCCO films were measured in Sherbrooke in a zero field and in a steady field of 16 T.

## VALUES OF $m^*$ AND $A_1$

<u>Hole-doped cuprates.</u> The values of $p$ and $m^*$ used in Fig. 3a are listed in Table 2 below. For Nd-LSCO, the value of $p$ with its error bar is taken from ref. 11. For LSCO, the value of $p$ is taken from refs. 33 and 34, and we assume the same error bar as for Nd-LSCO. The value of $m^*$ is obtained from the measured specific heat $\gamma$, via Eq. 2. For Nd-LSCO, the value of $\gamma$ with its error bar is taken from ref. 35, except for $p = 0.24$, where we take the average between the electronic specific heat $C_e / T$ at $T = 10$ K (12 mJ / K$^2$ mol) and at $T = 0.5$ K (22 mJ / K$^2$ mol), given that $C_e / T$ is not constant at low $T$ (ref. 35). For Bi2212 and LSCO, we estimate $\gamma$ and its error bar from the data published in ref. 32 and in refs. 33,34, respectively. With these values of $m^*$, we calculate $T_F = (\pi \hbar^2 / k_B)$ $(n d / m^*)$, using $n = (1-p) / (a^2 d)$ since the Fermi surface of Bi2212, LSCO and Nd-LSCO is electron-like at the dopings considered here (see Supplementary Section 1). We then obtain the Planckian limit on the resistivity slope, namely $A_1^\square = h / (2e^2 T_F)$, whose values are listed in the last column of Table 2 and plotted in Fig. 3b (open grey circles). For Bi2201, the values of $n$, $m^*$ and $A_1$ are given in Supplementary Section 10, with associated error bars and references.

**Table 2 | Effective mass and Planckian limit estimates in hole-doped cuprates.**

| Material | $p$ | $\gamma$ (mJ / K$^2$ mol) | $m^* / m_0$ | Ref. | $h / (2e^2 T_F)$ ($\Omega$ / K) |
|---|---|---|---|---|---|
| Bi2212 | 0.22 | 12 ± 2 | 8.4 ± 1.6 | 32 | 7.4 ± 1.4 |
| LSCO | 0.26 ± 0.005 | 14 ± 2 | 9.8 ± 1.7 | 34 | 8.9 ± 1.8 |
| | 0.29 ± 0.01 | 11 ± 1 | 7.7 ± 0.9 | 34 | 7.3 ± 1.2 |
| | 0.33 ± 0.01 | 6.9 ± 1 | 4.8 ± 0.8 | 33 | 4.9 ± 1.0 |
| Nd-LSCO | 0.24 ± 0.005 | 17 ± 5 | 12 ± 4 | 35 | 10.6 ± 3.7 |
| | 0.27 ± 0.01 | 11 ± 1 | 7.7 ± 0.9 | 35 | 7.1 ± 1.1 |
| | 0.36 ± 0.01 | 6.2 ± 1 | 4.3 ± 0.8 | 35 | 4.6 ± 1.0 |
| | 0.40 ± 0.01 | 5.4 ± 1 | 3.8 ± 0.8 | 35 | 4.2 ± 1.1 |

The values of $p$ and $A_1$ used in Fig. 3b are listed in Table 3 below. For Nd-LSCO, the value of $p$ with its error bar is taken from ref. 11. For LSCO, the value of $p$ is taken from refs. 8 and 40, and we assume the same error bar as for Nd-LSCO. For Nd-LSCO, the value of $A_1$ is obtained from a linear fit to the raw data in Fig. 1a ($p = 0.24$, at $H = 16$ T) and in Fig. 1c ($p = 0.22$ and 0.23, at $H = 33$ T and $P = 2$ GPa). Note that the MR is very weak in Nd-LSCO. For example, at $p = 0.24$, $A_1 = 0.47$ $\mu\Omega$ cm / K at $H = 33$ T (Fig. 1c) vs $A_1 = 0.49$ $\mu\Omega$ cm / K at $H = 16$ T (Fig. 1a). For LSCO, the value of $A_1$ is obtained from a linear fit to the raw data in Fig. 1b ($p = 0.26$, at $H = 18$ T) and to the MR-corrected data in Supplementary Section 7 ($p = 0.21$ and 0.23). Note that the MR in LSCO does not significantly change the slope $A_1$ (Fig. 1b vs Fig. S7). For Bi2212, the value of $A_1$ is obtained from a linear fit to MR-corrected data (Fig. 2b; see Supplementary Section 3). The error bar on $A_1$ is in all cases taken to be ± 10%, the estimated uncertainty in measuring the geometric factor of small samples. The values of $A_1$ listed in Table 3 are used to obtain the experimental values of $A_1^\square = A_1 / d$ that are plotted in Fig. 3b.

**Table 3 | Slope of *T*-linear resistivity in hole-doped cuprates.**

| Material | *p* | $A_1$ ($\mu\Omega$ cm / K) | *d* (Å) | $A_1^\square$ ($\Omega$ / K) | *H* (T) | Ref. |
|---|---|---|---|---|---|---|
| Bi2212 | 0.23 | 0.62 ± 0.06 | 7.73 ± 0.05 | 8.0 ± 0.9 | → 0 | this work |
| LSCO | 0.21 ± 0.005 | 1.0 ± 0.09 | 6.57 ± 0.05 | 15.2 ± 1.5 | → 0 | 8 |
| | 0.23 ± 0.005 | 0.75 ± 0.08 | 6.57 ± 0.05 | 11.4 ± 1.3 | → 0 | 8 |
| | 0.26 ± 0.005 | 0.54 ± 0.06 | 6.57 ± 0.05 | 8.2 ± 1.0 | 18 | 40 |
| Nd-LSCO | 0.22 ± 0.003 | 0.81 ± 0.08 | 6.64 ± 0.05 | 12.2 ± 1.3 | 33 | 12 |
| | 0.23 ± 0.003 | 0.68 ± 0.07 | 6.64 ± 0.05 | 10.2 ± 1.1 | 33 | 12 |
| | 0.24 ± 0.005 | 0.49 ± 0.05 | 6.64 ± 0.05 | 7.4 ± 0.8 | 16 | 7, 11 |

Electron-doped cuprates. The values of *x* and *m*\* used in Fig. 4a are listed in Table 4 below. For NCCO, the value of *x* is obtained from the frequency *F* of quantum oscillations, measured precisely (refs. 28, 29; see Supplementary Section 1), via $x = 1 - (2eFa^2 / h)$. The value of *m*\* is obtained directly from quantum oscillations, as reported (with error bar) in refs. 28, 29. For PCCO, *x* is taken to be the Cerium content, with an error bar ± 0.005. Here *m*\* is obtained from the measured specific heat γ, via Eq. 2, and the value of γ (with its error bar) is taken from ref. 30. With these values of *m*\*, we calculate $T_F = (\pi \hbar^2 / k_B) (n\, d / m^*)$, using $n = (1-x) / (a^2\, d)$ since the Fermi surface of NCCO and PCCO is hole-like (see Supplementary Section 1). We then obtain the Planckian limit on the resistivity slope, namely $A_1^\square = h / (2e^2\, T_F)$, whose values are listed in the last column of Table 4 and plotted in Fig. 4b (open grey circles).

**Table 4 | Effective mass and Planckian limit estimates in electron-doped cuprates.**

| Material | *x* | γ (mJ / K² mol) | *m*\* / $m_0$ | Ref. | $h / (2e^2\, T_F)$ ($\Omega$ / K) |
|---|---|---|---|---|---|
| PCCO | 0.15 ± 0.005 | 5.5 ± 0.3 | 3.6 ± 0.3 | 30 | 3.1 ± 0.3 |
| NCCO | 0.151 | - | 3.0 ± 0.3 | 28, 29 | 2.6 ± 0.3 |
| | 0.157 | - | 2.7 ± 0.1 | 28, 29 | 2.35 ± 0.1 |
| | 0.163 | - | 2.5 ± 0.1 | 28, 29 | 2.2 ± 0.1 |
| | 0.173 | - | 2.3 ± 0.05 | 28, 29 | 2.05 ± 0.05 |

The values of *x* and $A_1$ used in Fig. 4b are listed in Table 5 below. In all cases, *x* is taken to be the Cerium content, with an error bar ± 0.005. For PCCO at *x* = 0.17, the value of $A_1$ is obtained from a linear fit to the raw data in Supplementary Section 6. Within error bars, the same value is measured in all three PCCO films, whether at *H* = 0 or at *H* = 16 T. For LCCO, the value of $A_1$ is obtained from a linear fit to the raw data in Fig. 1d. Also listed in Table 5 are the values of $A_1$ obtained from a linear fit to the raw zero-field data in LCCO (see Supplementary Section 5). The error bar on $A_1$ is ± 15 % for our PCCO film, the uncertainty in measuring the film thickness. We apply the same error bar for LCCO. The values of $A_1$ listed in Table 5, both in zero field and in finite field, are used to obtain the experimental values of $A_1^\square = A_1 / d$ that are plotted in Fig. 4b (as open and closed squares, respectively).

**Table 5 | Slope of *T*-linear resistivity in electron-doped cuprates.**

| Material | $x$ | $A_1$ ($\mu\Omega$ cm / K) | $d$ (Å) | $A_1^{\square}$ ($\Omega$ / K) | $H$ (T) | Ref. |
|---|---|---|---|---|---|---|
| PCCO | $0.17 \pm 0.005$ | $0.10 \pm 0.015$ | $6.07 \pm 0.05$ | $1.7 \pm 0.3$ | 0 | this work |
| | | $0.10 \pm 0.015$ | $6.07 \pm 0.05$ | $1.7 \pm 0.3$ | 16 | this work |
| LCCO | $0.15 \pm 0.005$ | $0.18 \pm 0.03$ | $6.20 \pm 0.05$ | $3.0 \pm 0.45$ | 0 | 14 |
| | | $0.18 \pm 0.03$ | $6.20 \pm 0.05$ | $3.0 \pm 0.45$ | 8 | 14 |
| | $0.16 \pm 0.005$ | $0.145 \pm 0.02$ | $6.20 \pm 0.05$ | $2.4 \pm 0.35$ | 0 | 14 |
| | | $0.12 \pm 0.02$ | $6.20 \pm 0.05$ | $1.9 \pm 0.3$ | 6.5 | 14 |
| | $0.17 \pm 0.005$ | $0.10 \pm 0.015$ | $6.20 \pm 0.05$ | $1.7 \pm 0.3$ | 0 | 14 |
| | | $0.09 \pm 0.015$ | $6.20 \pm 0.05$ | $1.5 \pm 0.2$ | 4 | * |

* Courtesy of R. L. Greene.

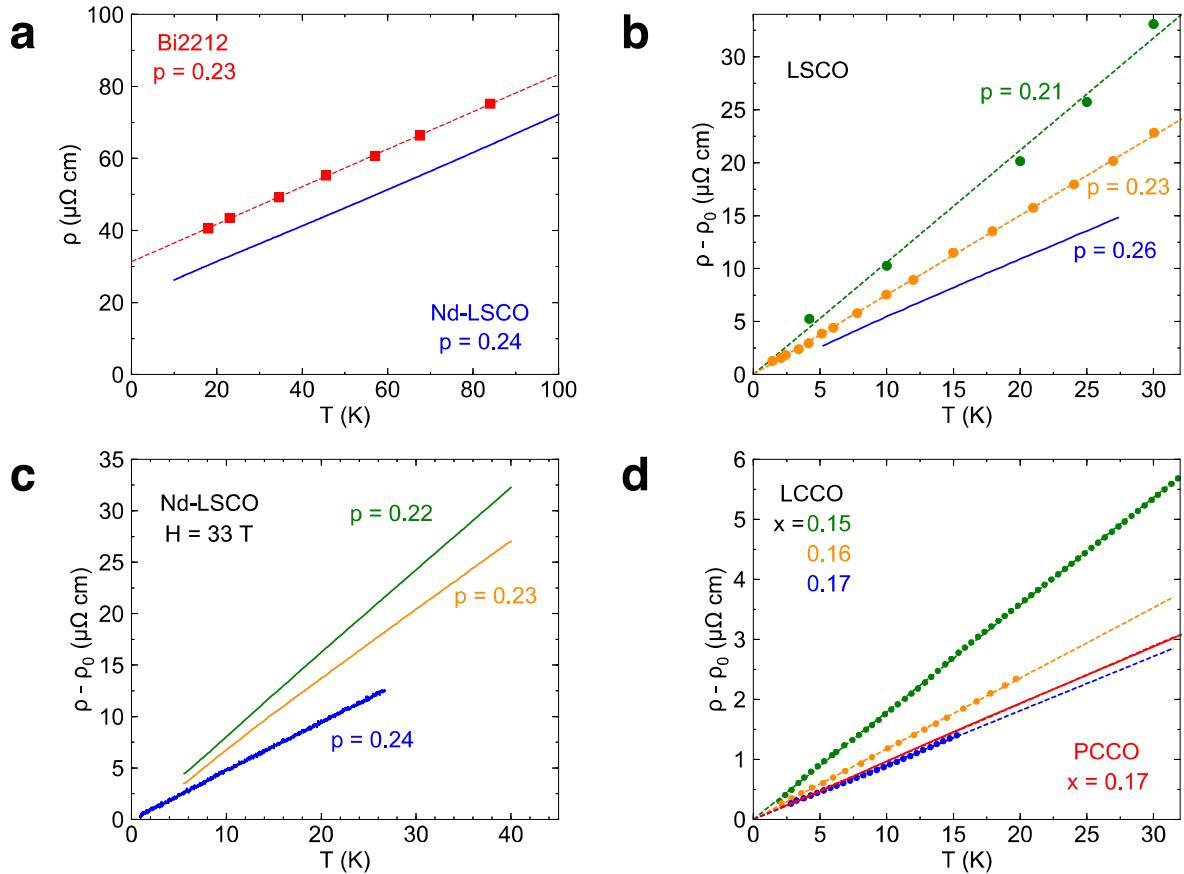

**Fig. 1 | *T*-linear resistivity in five overdoped cuprates.**

In-plane electrical resistivity ρ of cuprates showing a *T*-linear resistivity at low temperature. **a)** Nd-LSCO at *p* = 0.24 (blue, *H* = 16 T; from ref. 11) and Bi2212 at *p* = 0.23 (red squares, *H* = 55 T; this work, Fig. 2a). **b)** Temperature-dependent part of the resistivity, ρ(*T*) - ρ$_0$, for LSCO at *p* = 0.21 (green, *H* = 48 T; from ref. 8), *p* = 0.23 (orange, *H* = 48 T; from ref. 8), *p* = 0.26 (blue, *H* = 18 T; from ref. 40) (see Supplementary Section 7). **c)** ρ(T) - ρ$_0$ for Nd-LSCO at *H* = 33 T,  at *p* = 0.22 (green) and 0.23 (orange)  (from ref. 12) and at *p* = 0.24 (blue; from ref. 7). For *p* = 0.22 and 0.23, a pressure of 2 GPa was applied to suppress the pseudogap phase (see Supplementary Section 4). **d)** ρ(T) – ρ$_0$ for LCCO at *x* = 0.15 (green, *H* = 8 T), *x* = 0.16 (orange, *H* = 6.5 T) and *x* = 0.17 (blue, *H* = 4 T) (from ref. 14), and PCCO at *x* = 0.17 (red, *H* = 16 T; this work, see Supplementary Section 5). All dashed lines are a linear fit.

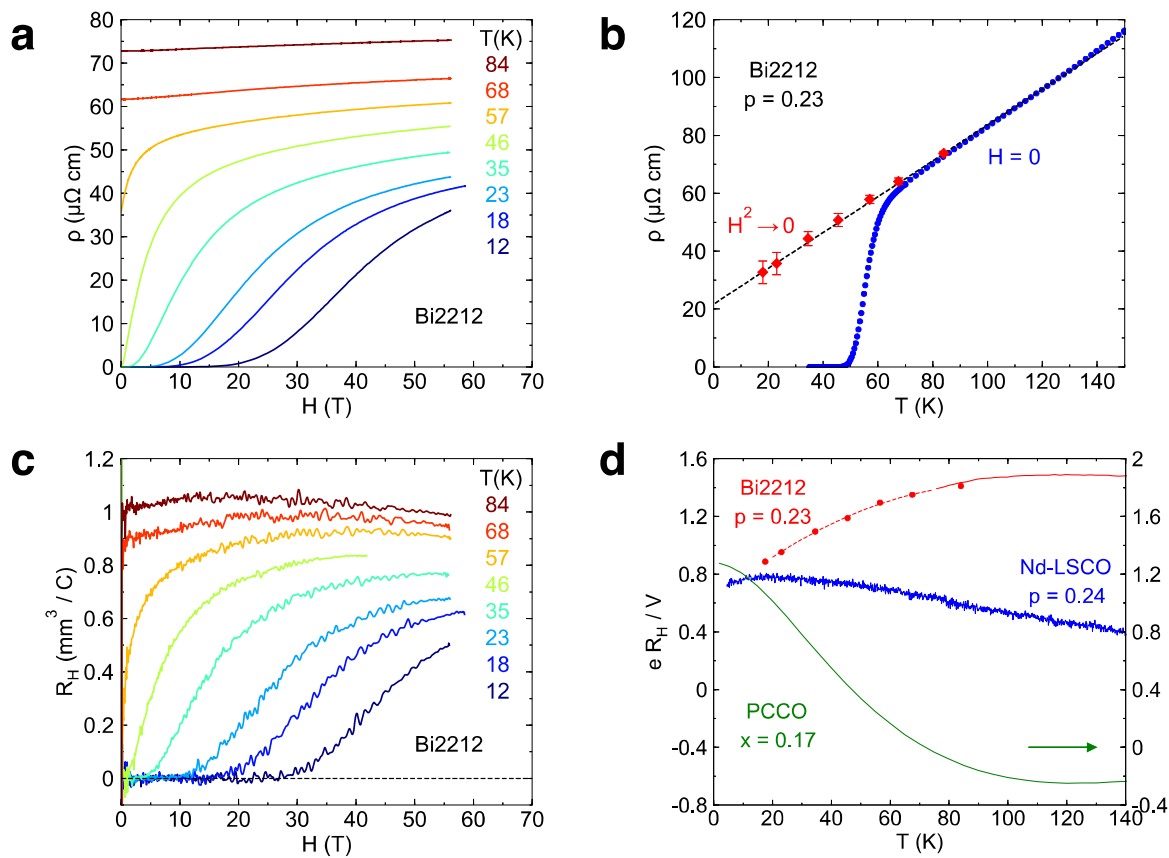

**Fig. 2 | Resistivity and Hall coefficient of our Bi2212 film.**

**a)** Resistivity of our Bi2212 film with $p = 0.23$ as a function of magnetic field, at temperatures as indicated. The value of $\rho$ at $H = 55$ T is plotted vs $T$ in Fig. 1a. **b)** Resistivity as a function of temperature, at $H = 0$ (blue). The red diamonds are high field data extrapolated to zero field by fitting $\rho(H)$ to a + b$H^2$ (see Methods and Supplementary Section 3). The line is a linear fit to the red diamonds. **c)** Hall coefficient of our Bi2212 film as a function of magnetic field, at temperatures as indicated. The value of $R_H$ at $H = 55$ T is plotted vs $T$ in (d). **d)** Hall coefficient as a function of temperature for three cuprates, plotted as $e R_H / V$, where $e$ is the electron charge and $V$ the volume per Cu atom: Bi2212 at $p = 0.23$ (red curve, $H = 9$ T; red dots, $H = 55$ T, panel c); Nd-LSCO at $p = 0.24$ (blue, $H = 16$ T; from ref. 11); PCCO at $x = 0.17$ (green, $H = 15$ T, right axis; from ref. 39). The dashed line is a guide to the eye.

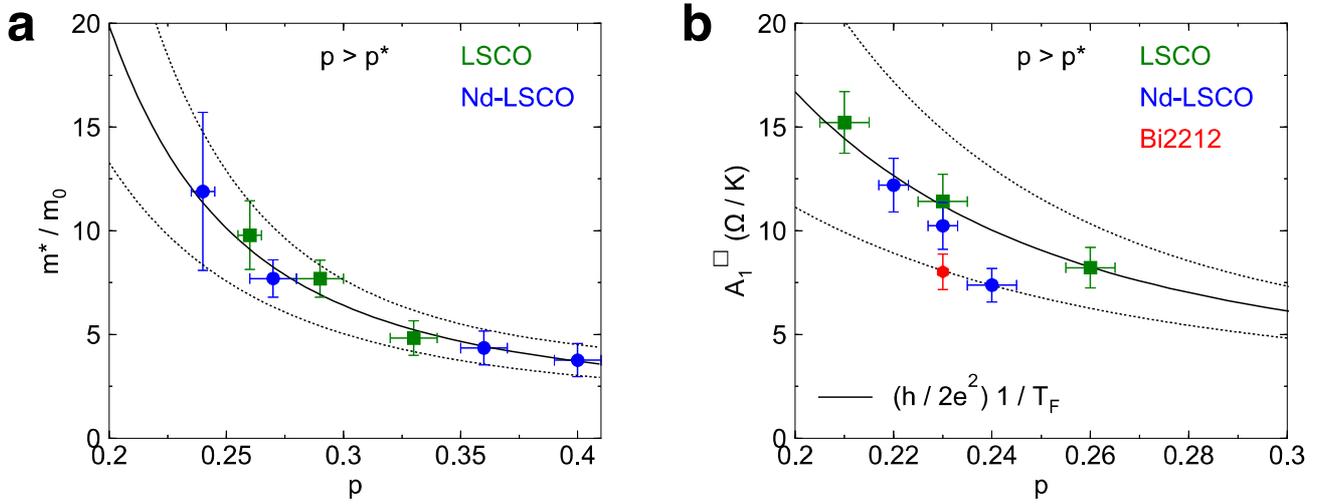

**Fig. 3 | Effective mass *m\** and slope of *T*-linear resistivity *$A_1^\square$* vs *p* in hole-doped cuprates.**

**a)** Effective mass *m\** of LSCO and Nd-LSCO as a function of doping, in units of the electron mass $m_0$, for $p > p^*$. The value of *m\** is obtained from specific heat data, via Eq. 2 (see Methods): in LSCO (green squares) at $p = 0.26$ and 0.29 (ref. 34), and at $p = 33$ (ref. 33); in Nd-LSCO (blue circles) at $p = 0.24$, 0.27, 0.36 and 0.40 (ref. 35). The solid line is a fit through the data, assumed to extend smoothly below $p = 0.24$. **b)** Slope of the *T*-linear resistivity, plotted as $A_1^\square = A_1 / d$, where *d* is the (average) distance between $CuO_2$ planes (see Methods): in LSCO (green squares) at $p = 0.21$, 0.23 and 0.26 (Supplementary Section 7); in Nd-LSCO (blue circles) at $p = 0.22$, 0.23 (Fig. 1c) and 0.24 (Fig. 1a); in Bi2212 at $p = 0.23$ (red hexagon, Fig. 2b). The experimental values are compared to the Planckian estimate (solid line) given by Eq. 1 with $\alpha = 1.0$, namely $A_1^\square = (m^* / n)\,(k_B / e^2 \hbar\, d)\,(= h / 2e^2 T_F)$, using *m\** from panel (a) (solid line in (a)) and $n = (1-p) / (a^2 d)$ (Methods). Error bars are explained in the Methods; the dotted lines represent the uncertainty on *m\**.

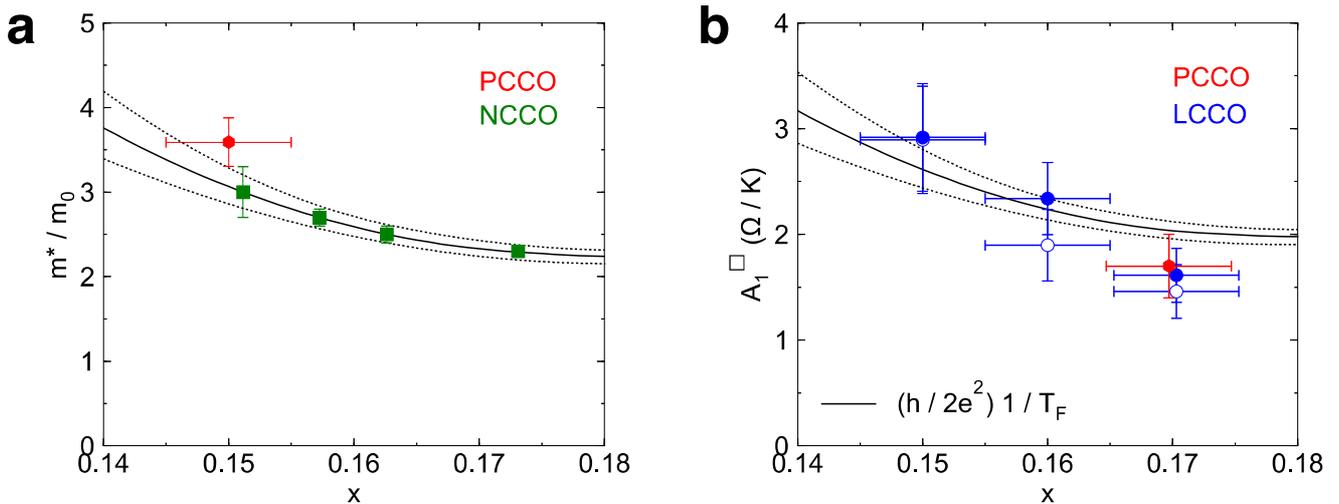

**Fig. 4 | Effective mass _m_\* and slope of _T_-linear resistivity _A_1□ vs _x_ in electron-doped cuprates.**

**a)** Effective mass $m^*$ of NCCO (green squares) as a function of doping, in units of $m_0$, measured by quantum oscillations[29]. Also shown is $m^*$ of PCCO at $x = 0.15$ (red hexagon), obtained from specific heat data[30] (see Methods). For NCCO, the value of $x$ is defined using the frequency $F$ of quantum oscillations, via $F = (1-x) \, (h/2ea^2)$. For PCCO, we use the nominal cerium content $x$, with its error bar (see Methods). The solid line is a fit through the NCCO data, assumed to extend smoothly below $x = 0.15$. **b)** Slope of the $T$-linear resistivity, plotted as $A_1^{\square} = A_1 / d$, where $d$ is the distance between $CuO_2$ planes (Methods): in LCCO (blue) at $x = 0.15$, 0.16 and 0.17, both in a field (open circles, Fig. 1d) and in zero field (full circles, Supplementary Section 5); in PCCO (red hexagon) at $x = 0.17$ (Supplementary Section 6). The experimental values are compared to the Planckian estimate (solid line) given by Eq. 1 with α = 1.0, namely $A_1^{\square} = (m^* / n) \, (k_B / e^2\hbar d)$, using $m^*$ from panel (a) (solid line in (a)) and $n = (1-x) / (a^2 d)$ (Methods). Error bars are explained in the Methods; the dotted lines represent the uncertainty on $m^*$.

| Material | | $n$ $(10^{27}\,\mathrm{m^{-3}})$ | $m^*$ $(m_0)$ | $A_1 / d$ $(\Omega / \mathrm{K})$ | $h / (2e^2\,T_\mathrm{F})$ $(\Omega / \mathrm{K})$ | $\alpha$ |
|---|---|---|---|---|---|---|
| Bi2212 | $p = 0.23$ | 6.8 | $8.4 \pm 1.6$ | $8.0 \pm 0.9$ | $7.4 \pm 1.4$ | $1.1 \pm 0.3$ |
| Bi2201 | $p \sim 0.4$ | 3.5 | $7 \pm 1.5$ | $8 \pm 2$ | $8 \pm 2$ | $1.0 \pm 0.4$ |
| LSCO | $p = 0.26$ | 7.8 | $9.8 \pm 1.7$ | $8.2 \pm 1.0$ | $8.9 \pm 1.8$ | $0.9 \pm 0.3$ |
| Nd-LSCO | $p = 0.24$ | 7.9 | $12 \pm 4$ | $7.4 \pm 0.8$ | $10.6 \pm 3.7$ | $0.7 \pm 0.4$ |
| PCCO | $x = 0.17$ | 8.8 | $2.4 \pm 0.1$ | $1.7 \pm 0.3$ | $2.1 \pm 0.1$ | $0.8 \pm 0.2$ |
| LCCO | $x = 0.15$ | 9.0 | $3.0 \pm 0.3$ | $3.0 \pm 0.45$ | $2.6 \pm 0.3$ | $1.2 \pm 0.3$ |
| TMTSF | $P = 11$ kbar | 1.4 | $1.15 \pm 0.2$ | $2.8 \pm 0.3$ | $2.8 \pm 0.4$ | $1.0 \pm 0.3$ |

**Table 1 | Slope of *T*-linear resistivity vs Planckian limit in seven materials.**

Comparison of the measured slope of the $T$-linear resistivity in the $T = 0$ limit, $A_1$, with the value predicted by the Planckian limit (Eq. 1; penultimate column), for four hole-doped cuprates (Bi2212, Bi2201, LSCO and Nd-LSCO), two electron-doped cuprates (PCCO and LCCO) and the organic conductor $(\mathrm{TMTSF})_2\mathrm{PF}_6$, as discussed in the text (and Supplementary Information). The ratio $\alpha$ of the experimental value, $A_1^\square = A_1 / d$, over the predicted value, is given in the last column. Although $A_1^\square$ varies by a factor 5, the ratio $m^* / n$ ($\sim 1/T_\mathrm{F}$) is seen to vary by the same amount, so that $\alpha = 1.0$ in all cases, within error bars.

[1] Löhneysen, H.v. *et al.* Non-Fermi-liquid behavior in a heavy-fermion alloy at a magnetic instability. *Phys. Rev. Lett.* **72**, 3262-3265 (1994).

[2] Fournier, P. *et al.* Insulator-metal crossover near optimal doping in $Pr_{2-x}Ce_xCuO_4$ : Anomalous normal-state low temperature resistivity. *Phys. Rev. Lett.* **81**, 4720-4723 (1998).

[3] Grigera, S. A. *et al.* Magnetic field-tuned quantum criticality in the metallic ruthenate $Sr_3Ru_2CuO_7$. *Science* **294,** 329-332 (2001).

[4] Doiron-Leyraud, N. *et al.* Correlation between linear resistivity and $T_c$ in the Bechgaard salts and the pnictide superconductor $Ba(Fe_{1-x}Co_x)_2As_2$. *Phys. Rev. B.* **80**, 214531 (2009).

[5] Coleman, P. & Schofield, A. J. Quantum criticality. *Nature* **433**, 226-229 (2005).

[6] Martin, S. *et al.* Normal-state transport properties of $Bi_{2+x}Sr_{2-y}CuO_{6+\delta}$ single crystals. *Phys. Rev. B* **41**, 846-849 (1990).

[7] Daou, R. *et al.* Linear temperature dependence of resistivity and change in the Fermi surface at the pseudogap critical point of a high-$T_c$ superconductor. *Nat. Phys.* **5**, 31-34 (2009).

[8] Cooper, R. A. *et al.* Anomalous criticality in the electrical resistivity of $La_{2-x}Sr_xCuO_4$. *Science* **323**, 603-607 (2009).

[9] Zaanen, J. Why the temperature is high. *Nature* **430**, 512-513 (2004).

[10] Bruin, J. A. N. *et al.* Similarity of scattering rates in metals showing *T*-linear resistivity. *Science* **339**, 804-807 (2013).

[11] Collignon, C. *et al.* Fermi-surface transformation across the pseudogap critical point of the cuprate superconductor $La_{1.6-x}Nd_{0.4}Sr_xCuO_4$. *Phys. Rev. B.* **95**, 224517 (2017).

[12] Doiron-Leyraud, N. *et al.* Pseudogap phase of cuprate superconductors confined by Fermi surface topology. *Nat. Commun.* **8**, 2044 (2017).

[13] Jin, K. *et al.* Link between spin fluctuations and electron pairing in copper oxide superconductors. *Nature* **476**, 73-75 (2011).

[14] Sarkar, T. *et al.* Fermi surface reconstruction and anomalous low-temperature resistivity in electron-doped $La_{2-x}Ce_xCuO_4$. *Phys. Rev. B.* **96**, 155449 (2017).

[15] Löhneysen, H.v. *et al.* Fermi-liquid instabilities at magnetic quantum phase transitions. *Rev. Mod. Phys.* **79**, 1015-1075 (2007).

[16] Dagan, Y. *et al.* Evidence for a quantum phase transition in $Pr_{2-x}Ce_xCuO_4$ from transport measurements. *Phys. Rev. Lett.* **92**, 167001 (2004).

[17] Tafti, F. F. *et al.* Nernst effect in the electron-doped cuprate superconductor $Pr_{2-x}Ce_xCuO_4$: Superconducting fluctuations, upper critical field $H_{c2}$, and the origin of the $T_c$ dome. *Phys. Rev. B.* **90**, 024519 (2014).

[18] Motoyama, E. M. *et al.* Spin correlations in the electron-doped high-transition-temperature superconductor $Nd_{2-x}Ce_xCuO_4$. *Nature* **445**, 186-189 (2007).

[19] Matt, Ch. *et al.* Electron scattering, charge order, and pseudogap physics in $La_{1.6-x}Nd_{0.4}Sr_xCuO_4$: An angle-resolved photoemission spectroscopy study. *Phys. Rev. B* **92**, 134524 (2015).

[20] Yoshida, T. *et al.* Systematic doping evolution of the underlying Fermi surface of $La_{2-x}Sr_xCuO_4$. *Phys. Rev. B* **74**, 224510 (2006).

[21] Kaminski, A. *et al.* Change of Fermi-surface topology in $Bi_2Sr_2CaCu_2O_{8+\square}$ with doping. *Phys. Rev. B* **73**, 174511 (2006).

[22] Benhabib, S. *et al.* Collapse of the normal-state pseudogap at a Lifshitz transition in the $Bi_2Sr_2CaCu_2O_{8+\square}$ cuprate superconductor. *Phys. Rev. Lett.* **114**, 147001 (2015).

[23] Hussey, N. E. *et al.* Phenomenology of the normal-state in-plane transport properties of high-$T_c$ cuprates. *J. Phys.: Condens. Matter* **20**, 123201 (2008).

[24] Fauqué, B. *et al.* Dispersion of the odd magnetic resonant mode in near-optimally doped $Bi_2Sr_2CaCu_2O_{8+\square}$. *Phys. Rev. B* **76**, 214512 (2007).

[25] Tranquada, J. M. *et al.* Coexistence of, and competition between, superconductivity and charge-stripe order in $La_{1.62-x}Nd_{0.4}Sr_xCuO_4$. *Phys. Rev. Lett.* **78**, 338 (1997).

[26] Hussey, N. E. *et al.* Generic strange metal behavior of overdoped cuprates. *J. Phys.: Conf. Series* **449**, 012004 (2013).

[27] Matsui, H. *et al.* Evolution of the pseudogap across the magnet-superconductor phase boundary of $Nd_{2-x}Ce_xCuO_4$. *Phys. Rev. B* **75**, 224514 (2007).

[28] Helm, T. *et al.* Evolution of the Fermi surface of the electron-doped high-temperature superconductor $Nd_{2-x}Ce_xCuO_4$ revealed by Shubnikov–de Haas oscillations. *Phys. Rev. Lett.* **103**, 157002 (2009).

[29] Helm, T. *et al.* Correlation between Fermi surface transformations and superconductivity in the electron-doped high-$T_c$ superconductor $Nd_{2-x}Ce_xCuO_4$. *Phys. Rev. B* **92**, 094501 (2015).

[30] Yu, W. *et al*. Magnetic-field dependence of the low-temperature specific heat of the electron-doped superconductor $Pr_{1.85}Ce_{0.15}CuO_4$. *Phys. Rev. B.* **72**, 212512 (2005).

[31] Bangura, A. *et al*. Fermi surface and electronic homogeneity of the overdoped cuprate superconductor $Tl_2Ba_2CuO_{6+\delta}$ as revealed by quantum oscillations. *Phys. Rev. B.* **82**, 140501R (2010).

[32] Loram, J. W. *et al*. Evidence on the pseudogap and the condensate from the electronic specific heat. *J. Phys. Chem. Solids* **62**, 59-64 (2001).

[33] Nakamae, S. *et al*. Electronic ground state of heavily overdoped nonsuperconducting $La_{2-x}Sr_xCuO_4$. *Phys. Rev. B.* **68**, 100502R (2003).

[34] Wang, Y. *et al*. Weak-coupling *d*-wave BCS superconductivity and unpaired electrons in overdoped $La_{2-x}Sr_xCuO_4$ single crystals. *Phys. Rev. B.* **76**, 064512 (2007).

[35] Michon, B. *et al*. Thermodynamic signatures of quantum criticality in cuprate superconductors. arXiv:1804.08502 (2018).

[36] Davison, R. A., Schalm, K. & Zaanen, J. Holographic duality and the resistivity of strange metals. *Phys. Rev. B.* **89**, 245116 (2014).

[37] Hartnoll, S. A. Theory of universal incoherent metallic transport. *Nat. Phys.* **11**, 54-61 (2015).

[38] Song, X. Y. *et al*. Strongly correlated metal built from Sachdev-Ye-Kitaev models. *Phys. Rev. Lett.* **119**, 216601 (2017).

[39] Charpentier, S. *et al*. Antiferromagnetic fluctuations and the Hall effect of electron-doped cuprates: possibility of a quantum phase transition at underdoping. *Phys. Rev. B.* **81**, 104519 (2010).

[40] Hussey, N. E. *et al*. Dichotomy in the *T*-linear resistivity in hole-doped cuprates. *Phil. Trans. R. Soc. A* **369**, 1626-1639 (2011).

# SUPPLEMENTARY INFORMATION

**CONTENTS**

**Section 1**

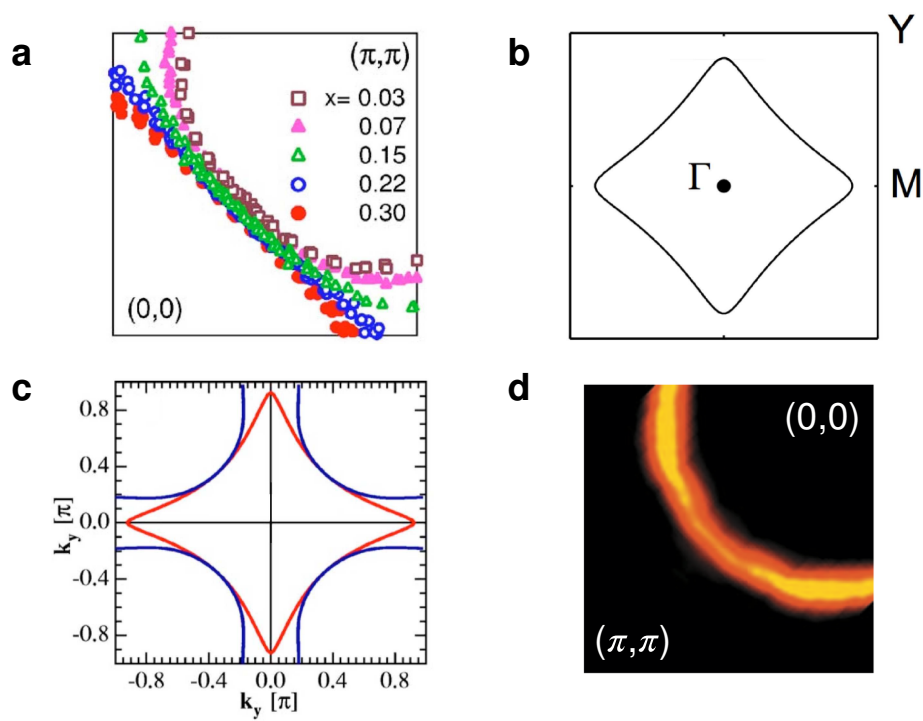

**Figure S1 | Fermi surface of the four cuprates.**

Fermi surface of four different cuprates, as measured by ARPES: **a)** LSCO at four dopings as indicated (from ref. 20); **b)** Nd-LSCO at *p* = 0.24 (from ref. 19); **c)** Bi2212 at *p* = 0.23 (from ref. 21); **d)** NCCO at *x* = 0.17 (from ref. 27). Note that all are single-layer materials and so have only a single Fermi surface, except for Bi2212, which is a bi-layer material, with two Fermi surfaces, one of which is hole-like (blue), the other electron-like (red).

The Fermi surface area of NCCO (Fig. S1d) is known precisely from the frequency $F$ of quantum oscillations. For the following nominal $x$ values, the following values of $F$ and associated $m^*$ were measured [41]: $x$ = 0.15, 0.16, 0.165, 0.17; $F$ = 10.96 ± 50, 11.10 ± 50, 11.17 ± 100, 11.25 ± 100 kT; $m^*$ = 3.0 ± 0.3, 2.7 ± 0.1, 2.5 ± 0.1, 2.3 ± 0.05. The precise values of $x$ obtained from the measured $F$ via the Luttinger rule, $x = 1 - (2eFa^2/h)$, are listed in Table 4 of the Methods.

## Section 2

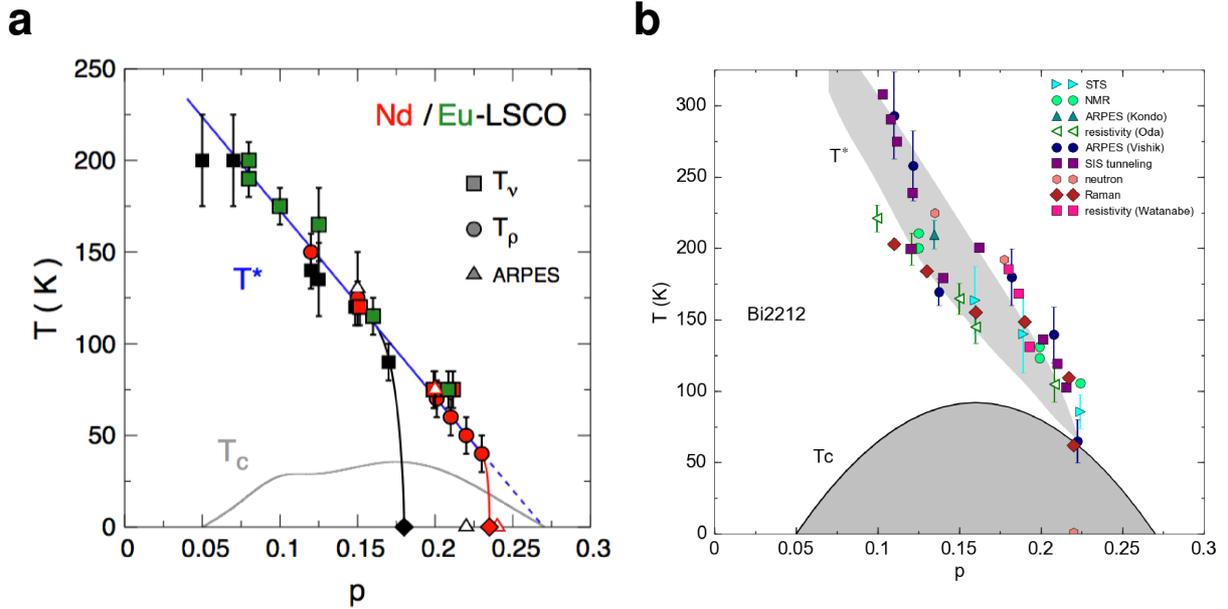

**Figure S2 | Phase diagram of hole-doped cuprates.**

Temperature-doping phase diagrams: **(a)** Nd-LSCO (red) and LSCO (black) (from ref. 42); **(b)** Bi2212 (adapted from ref. 43). The pseudogap phase ends at the critical doping $p^*$ = 0.23 in Nd-LSCO, $p^*$ = 0.18-0.19 in LSCO, and $p^*$ = 0.22 in Bi2212.

## Section 3

In Fig. 2a, isotherms in Bi2212 exhibit a small normal-state magnetoresistance (MR). In Fig. S3a, we see that this MR grows as $H^2$, at $T$ = 84 K. To correct for the MR at lower $T$, we fit the data to $\rho(H) = \rho(H^2 \rightarrow 0) + cH^2$ above a threshold field (dashed lines in Fig. S3a), namely : 40 T for $T$ = 68, 57 and 46 K; 50 T for $T$ = 35 and 23 K; 55T for $T$ = 18 K. In Fig. S3b, we plot $\rho(H^2 \rightarrow 0)$ vs $T$ (red diamonds) and observe that $\rho(H^2 \rightarrow 0)$ is the linear continuation (dashed line) of the $H$ = 0 data at high $T$ (black dots), within error bars. This shows that in the absence of MR, the normal-state resistivity of Bi2212 is $T$-linear from $T \sim$ 120 K down to at least $T$ = 18 K. The slope of $\rho(H^2 \rightarrow 0)$ vs $T$ (red diamonds, Figs. 2b and S3b) is $A_1$ = 0.62 μΩ cm / K (Table 2, Methods), while the slope of $\rho(H$=55T) vs $T$ (red squares, Fig. 1a) is $A_1$ = 0.50 μΩ cm / K. Note that the same approach was used to correct for the MR in LSCO (see ref. 8).

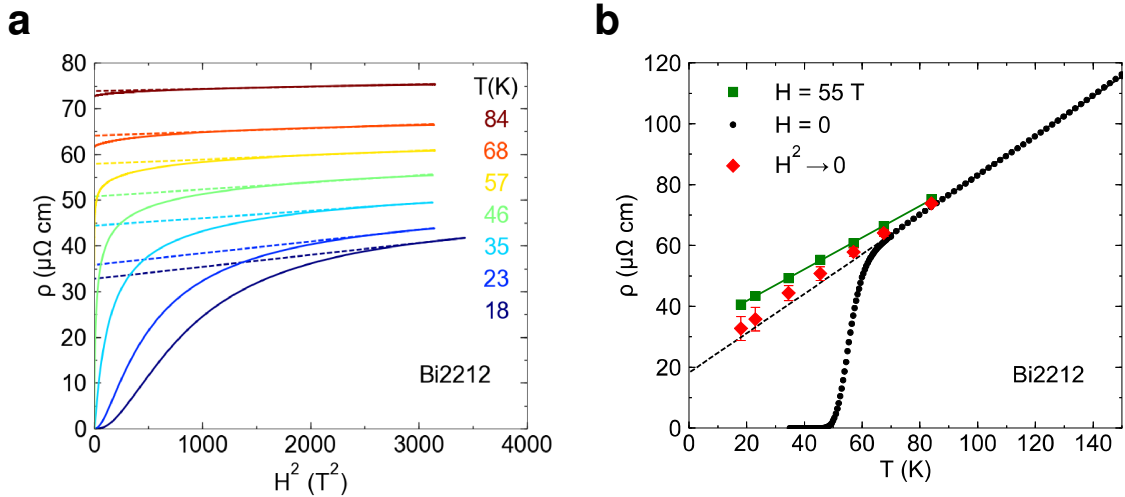

**Figure S3 | Correcting for the magneto-resistance in Bi2212.**

**a)** Magnetic field dependence of the resistivity ρ in our sample of Bi2212, plotted vs $H^2$, at different temperatures as indicated. The dashed lines are linear fits to the data at high $H$, *i.e.* $\rho(H) = \rho(H^2{\to}0) + cH^2$. **b)** Temperature dependence of ρ : at $H = 0$ (black dots), at $H = 55$ T (green squares), and $\rho(H^2{\to}0)$ (red diamonds) obtained from the fits in panel a). The green line is a linear fit to ρ(55T); the dashed black line a linear fit to the $H = 0$ data between 80 K and 130 K.

## Section 4

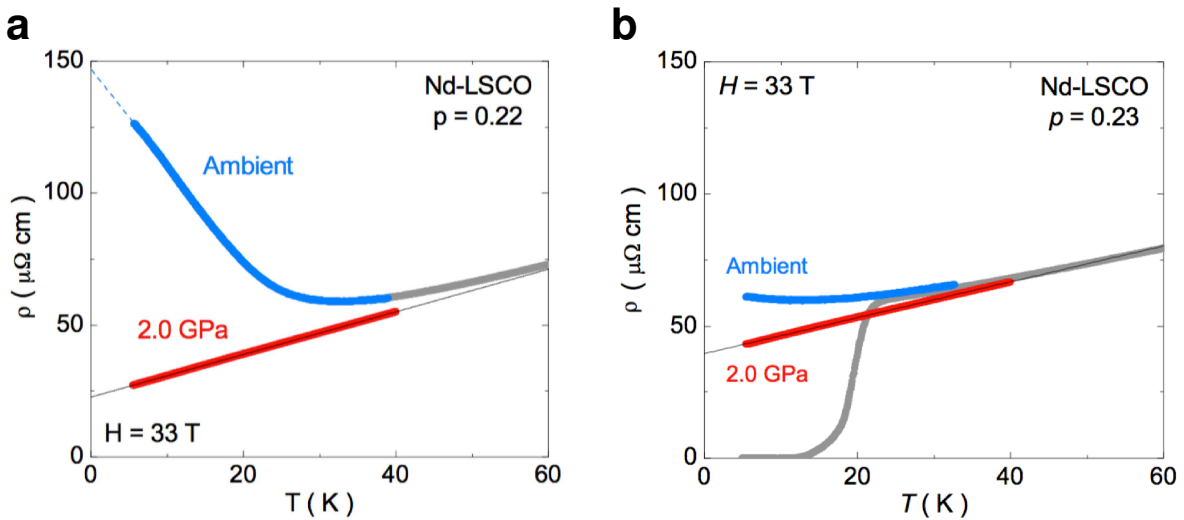

**Figure S4 | Resistivity of Nd-LSCO under pressure.**

Normal-state resistivity of Nd-LSCO at $p = 0.22$ (**a**) and $p = 0.23$ (**b**), measured at $H = 0$ (grey) and $H = 33$ T at ambient pressure (blue) and at $P = 2.0$ GPa (red) (from ref. 12). The effect of pressure is to suppress the pseudogap phase, by moving $p^*$ below 0.22. This shows that the resistivity is then perfectly linear at low $T$.

*T*-linear resistivity in Nd-LSCO was first reported in 2009, at $p$ = 0.24 (ref. 7). At lower doping, the resistivity shows an upturn at low $T$, the signature of the pseudogap (refs. 7, 11). This yields $p^*$ = 0.23 in Nd-LSCO (ref. 11), consistent with ARPES data that find the pseudogap in Nd-LSCO to close at a doping above $p$ = 0.20 and below $p$ = 0.24 (ref. 19).

It was recently found that $p^*$ can be lowered by the application of hydrostatic pressure (ref. 12). A pressure of 2 GPa moves $p^*$ below 0.22, *i.e.* it removes the resistivity upturn in Nd-LSCO at $p$ = 0.22 and $p$ = 0.23 (Fig. S4). Having removed the pseudogap, one finds a perfectly linear $T$ dependence as $T \rightarrow 0$ (Fig. S4). We then see that the regime of $T$-linear resistivity is stretched from $p$ = 0.24 down to $p^*$, producing an anomalous range similar to that found in LSCO (ref. 8). In that range, we again observe that $A_1$ increases with decreasing $p$ (Figs. 1c and 3b).

## Section 5

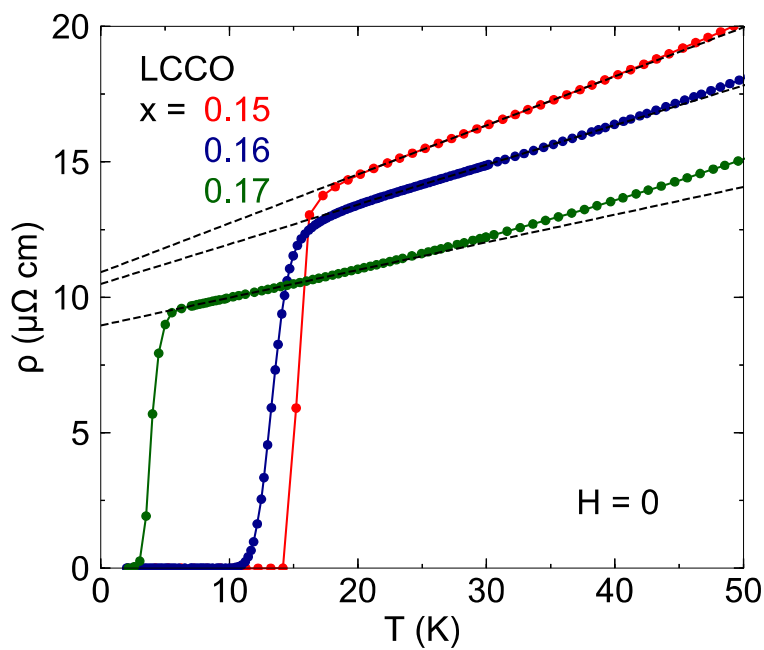

**Figure S5 | Resistivity of LCCO films.**

Temperature dependence of the resistivity in LCCO in zero field at three dopings, as indicated (from ref. 14, and courtesy of R.L. Greene). Lines are a linear fit to low-$T$ data.

In electron-doped cuprates, $T$-linear resistivity was first observed in PCCO at $x$ = 0.17 in 1998 (ref. 2). At the time, thin films contained traces of an extra phase, and so the absolute value of the resistivity was not reliable. Since 2009 (ref. 39), this has been resolved. In recent measurements on PCCO (refs. 17, 39) and on LCCO (refs. 13,14), a $T$-linear resistivity at low $T$ with reliable absolute value has been reported, giving $A_1$ = 0.1 μΩ cm / K in both PCCO and LCCO at $x$ = 0.17.

In Fig. S5, we reproduce the zero-field resistivity of LCCO at $x$ = 0.15, 0.16 and 0.17, from ref. 14 (and courtesy of R. L. Greene). Linear fits at low $T$ yield the values of $A_1$ listed in Table 5 of Methods, which give $A_1^{\square}$ = 3.0, 2.4 and 1.7 Ω / K at $x$ = 0.15, 0.16 and 0.17, respectively.

In Fig. 1d, we reproduce the in-field resistivity of LCCO at $x$ = 0.15, 0.16 and 0.17, from ref. 14 (and courtesy of R. L. Greene). Linear fits at low $T$ yield values of $A_1$ that are very similar to the zero-field values (see Table 5 in Methods).

## Section 6

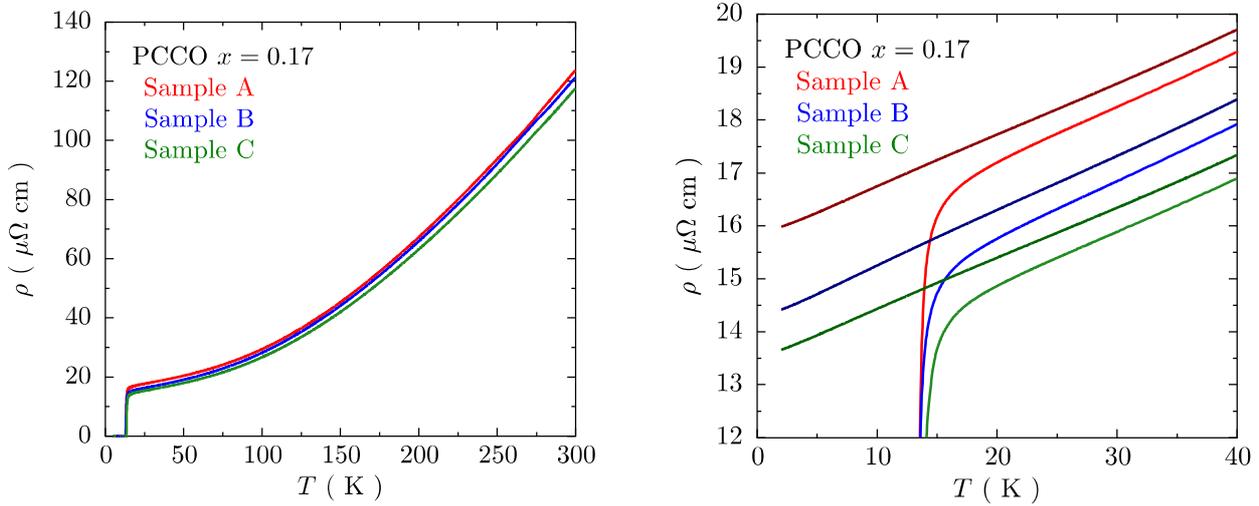

**Figure S6 | Resistivity of PCCO films.**

**a)** Temperature dependence of the resistivity in our three PCCO films with $x$ = 0.17, in zero field. **b)** Zoom on the low-temperature regime, at $H$ = 0 (pale) and $H$ = 16 T (dark). The 16 T curve for sample C is also shown in Fig. 1d.

To double-check the value of $A_1$ in PCCO at $x$ = 0.17, we have grown and measured three films of PCCO at $x$ = 0.17, with $T_c$ = 13.1 K (sample A), 13.0 K (sample B) and 13.4 K (sample C). These films have a very similar residual resistivity ratio, $RRR$ = $\rho$(300K)/$\rho$($T \rightarrow 0$) = 8.2, 8.8 and 9.1, respectively. The sample thickness $t$ = 230 ± 30 um is measured by the width of the x-ray diffraction peak. For films of that thickness, the uncertainty is roughly ± 15%. As shown in Fig. S6, we obtain $A_1$ = 0.10 μΩ cm / K on all three films (at $H$ = 0), in good agreement with published data. Applying a field of 16 T suppresses superconductivity completely ($H_{c2}$ = 3 T; ref. 17) and extends the linear $T$ dependence to the lowest $T$. The slope at $H$ = 16 T is the same as in zero field (see Table 5 in Methods). We conclude that $A_1^{\square}$= 1.7 ± 0.3 Ω / K in PCCO at $x$ = 0.17 (Table 1).

## Section 7

The low-$T$ resistivity of LSCO was measured by Cooper *et al*. from $p$ = 0.18 up to $p$ = 0.33, by applying a magnetic field up to 60 T (ref. 8). At $p$ = 0.21, 0.23 and 0.26, 48 T is sufficient to suppress superconductivity down to (at least) 2 K. At those three dopings, the resistivity is linear as $T \rightarrow 0$ , below a certain temperature $T_0$ . At $p$ = 0.23, for example, a perfect linearity is observed in the raw data at 48 T below 50 K (down to at least 2 K). The slope $A_1$ in 48 T is the same as the slope in zero field observed between $T_c$ and $T_0 \sim 75$ K. At $p = 0.21$, $T_0 \sim 150$ K, while at $p = 0.26$, $T_0 \sim 30$ K (ref. 40). The value of $A_1$ increases with decreasing $p$ (Fig. 1b, Fig. 3b). At $p > 0.26$, the resistivity is no longer purely $T$-linear at low $T$. Instead, it can be fit to $A_1 T + A_2 T^2$ at $p = 0.29$ and to $A_2 T^2$ at $p = 0.33$ (i.e. $A_1 = 0$). So the $T$-linear resistivity as $T \to 0$ is observed in LSCO from $p = 0.26$ down to at least $p = 0.21$, possibly down to $p = 0.18$ (where it is more difficult to suppress superconductivity), i.e. down to $p^* \sim 0.18$-$0.19$. In LSCO, $p^*$ is identified as the doping below which the resistivity is no longer $T$-linear at low $T$, and $p^* = 0.18$-$0.19$ is consistent with ARPES data that find the pseudogap in LSCO to close above $p = 0.15$ and below $p = 0.22$ (ref. 20). The fact that $T$-linear resistivity is observed over a sizable range of doping is considered anomalous and requires an explanation.

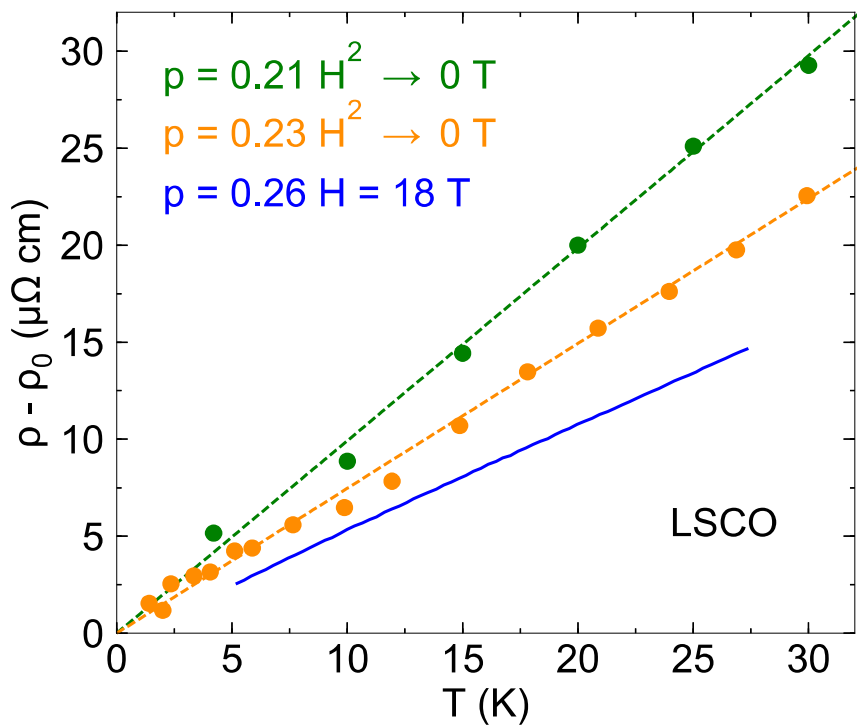

**Figure S7 | Resistivity of LSCO corrected for MR.**

Temperature-dependent part of the normal-state resistivity of LSCO, $\rho(T) - \rho_0$, at $p = 0.21$ (green) and $p = 0.23$ (orange), from ref. 8, and at $p = 0.26$ (blue, $H = 18$ T; from ref. 40). The green and blue dots are the MR-corrected resistivity, $\rho(H^2 \to 0)$, obtained in ref. 8 from a fit of $\rho$ vs $H$ isotherms to $\rho(H) = \rho(H^2 \to 0) + cH^2$. The green and blue lines are a linear fit to $\rho(H^2 \to 0)$ vs $T$, whose slope $A_1$ is given in Table 3 of the Methods.

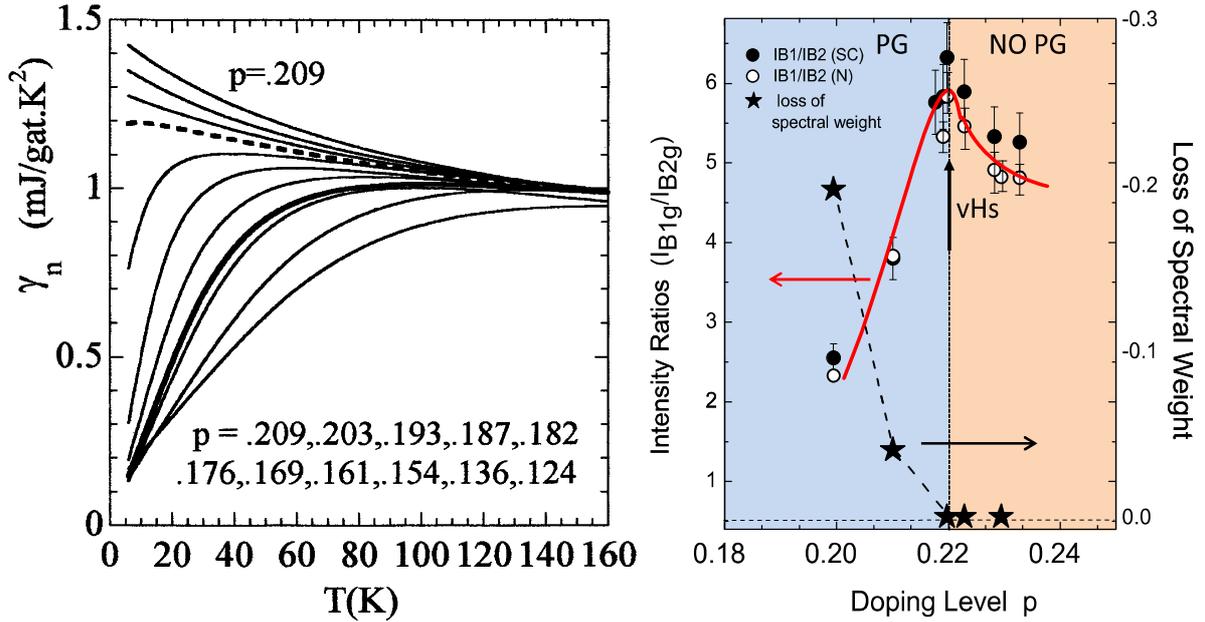

**Figure S8 | Specific heat and Raman intensity in Bi2212.**

**a)** Normal-state specific heat coefficient γ of Bi2212 vs temperature, at various dopings as indicated, estimated from an analysis of data up to high temperature (from ref. 32). **b)** Ratio of Raman intensities in Bi2212 vs doping, for the modes that select anti-nodal ($B_{1g}$) vs nodal ($B_{2g}$) regions in $k$-space (from ref. 22).

In the $T$ = 0 limit, γ in Bi2212 is seen to increase from 1.2 at $p$ = 0.187 to 1.5 mJ / gat. K$^2$ at $p$ = 0.209. A linear extrapolation up to $p$ = 0.22 yields γ = 1.65 ± 0.15 mJ / gat. K$^2$ at $p$ = 0.22, which converts to γ = 12 ± 2 mJ / K$^2$ mol-Cu (Table 2, Methods). The peak in the Raman intensity ratio, which is sensitive to the opening of the anti-nodal pseudogap (PG), shows that the pseudogap critical point in Bi2212 is $p^*$ = 0.22. Our sample has a doping of $p$ = 0.23, and so is very slightly above $p^*$. It is reasonable to assume that γ at $T$ = 0 (panel a) will continue to increase until $p$ reaches $p^*$.

**Section 9**

The organic conductor (TMTSF)$_2$PF$_6$ is a well-characterized single-band metal. When tuned to its QCP (by pressure), (TMTSF)$_2$PF$_6$ exhibits a resistivity that is perfectly $T$-linear below 8 K, down to the lowest measured temperature (~ 0.1 K), with a slope $A_1$ = 0.38 ± 0.04 μΩ cm / K (ref. 4). With a carrier density $n$ = 1.4×10$^{27}$ m$^{-3}$ (ref. 44) and an effective mass $m^*$ = 1.0 − 1.3 $m_0$ (ref. 45), we get $A_1$ = α ($m^*$ / $n$) ($k_B$ / $e^2 \hbar$) = α (0.33 − 0.43 μΩ cm / K), so that α = 1.0 ± 0.3.

To calculate the 2D sheet resistance listed in Table 1, we divide $A_1$ by the interlayer separation along the $c$ axis, $d$ = 1.35 nm, yielding $A_1^\square = A_1 / d$ = 2.8 ± 0.3 Ω / K.

## Section 10

In the single-layer cuprate Bi2201, the pseudogap critical point is located at very high doping, near the end of the superconducting dome, namely where $T_c$ ~ 10 K [46]. The Fermi surface measured by ARPES is also found to change topology from hole-like to electron-like near the end of the superconducting dome [47]. The volume of the Fermi surface at that doping is such that $p$ ~ 0.4 [47], so that the carrier density contained in the electron-like Fermi surface is $n = 1 - p$ ~ 0.6.

Near the end of the superconducting dome, at $T_c$ ~ 7 K, the resistivity is found to be perfectly $T$-linear [6]. In two crystals with nearly the same doping ($T_c$), $A_1$ = 0.74 and 1.06 $\mu\Omega$ cm / K [6]. Taking the average of those two values, consistent with typical error bars on geometric factors (±15%), we get $A_1$ = 0.9 ± 0.2 $\mu\Omega$ cm / K. Dividing by the interlayer spacing, which is two times larger in Bi2201 than in LSCO, we get $A_1^\square$ = 8 ± 2 $\Omega$ / K. Remarkably, this is the same value, within error bars, as in Bi2212 and Nd-LSCO, all at their respective critical dopings, namely $p^*$ = 0.22, 0.23, and 0.4 (Table 1).

We can estimate $m^*$ from specific heat data measured on a Bi2201 crystal with $T_c$ = 19 K [48], at a doping slightly below $p^*$ [46]. With increasing field to suppress superconductivity, $\gamma$ increases from 6 mJ / K$^2$ mol at $H = 0$ to 8 mJ / K$^2$ mol at $H = 6$ T, and is estimated to reach 10 mJ / K$^2$ mol at the critical field $H_{c2}$ = 18 T [48]. Given the uncertainty in the latter estimation, we take $\gamma$ = 10 ± 2 mJ / K$^2$ mol , which yields $m^*$ = 7 ± 1.5 $m_0$. Note that $\gamma$ may be somewhat larger at the slightly higher doping ($p$ ~ $p^*$) where $T$-linear resistivity was measured (see Supplementary Section 8 for a similar situation with respect to the specific heat data in Bi2212.)

Using $n$ = 0.6 and $m^*$ = 7 ± 1.5 $m_0$ , we calculate the value predicted for the Planckian limit: $A_1^\square = (m^* / n d) (k_B / e^2 \hbar)$ = 8 ± 2 $\Omega$ / K. The ratio of experimentally measured to theoretically predicted values of $A_1^\square$ is therefore $\alpha$ = 1.0 ± 0.4 (Table 1).

## SUPPLEMENTARY REFERENCES

[41] Helm, T. Electronic properties of electron-doped cuprate superconductors probed by high-field magneto-transport. *PhD thesis*. Technical University, Munich (2013).

[42] Cyr-Choinière, O. *et al.* Pseudogap temperature $T^*$ of cuprates from the Nernst effect. *Phys. Rev. B* **97**, 064502 (2018).

[43] Vishik, I. M. *et al.* Phase competition in trisected superconducting dome. *PNAS* **109**, 18332-18337 (2012).

[44] Moser, J. *et al.* Hall effect in the normal phase of the organic superconductor (TMTSF)$_2$PF$_6$. *Phys. Rev. Lett.* **84**, 2674-2677 (2000).

[45] Uji, S. *et al.* Rapid oscillations and Fermi-surface reconstruction due to spin-density-wave formation in the organic conductor (TMTSF)$_2$PF$_6$. *Phys. Rev. B* **55**, 12446-12453 (1997).

[46] Kawasaki, S. *et al.* Carrier-concentration dependence of the pseudogap ground state of superconducting Bi$_2$Sr$_{2-x}$La$_x$CuO$_{6+\delta}$ revealed by $^{63,65}$Cu-nuclear magnetic resonance· in very high magnetic fields. *Phys. Rev. Lett.* **105**, 137002 (2010).

[47] Kondo, T. *et al.* Hole-concentration dependence of band structure in (Bi,Pb)$_2$(Sr,La)$_2$CuO$_{6+\delta}$ determined by the angle-resolved photoemission spectroscopy. *J. Electron Spectroscopy and Related Phenomena* **137-140**, 663-668 (2004).

[48] Ikuta, H. *et al.* Low-temperature specific heat of overdoped Bi2201 single crystals. *Physica C* **388-389**, 361-362 (2003).